# AMIEDoT: An annotation model for document tracking and recommendation service

ROBERT Charles Abiodun

LORIA- Campus Scientifique, B. P. 239, 54506

Vandoeuvre-Lès-Nancy, France

Tel : +33383593000

Email : abiodun-charles.robert@loria.fr

**ABSTRACT**

The primary objective of document annotation in whatever form, manual or electronic is to allow those who may not have control to original document to provide personal view on information source. Beyond providing personal assessment to original information sources, we are looking at a situation where annotation made can be used as additional source of information for document tracking and recommendation service. Most of the annotation tools existing today were conceived for their independent use with no reference to the creator of the annotation. We propose AMIEDoT (Annotation Model for Information Exchange and Document Tracking) an annotation model that can assist in document tracking and recommendation service. The model is based on three parameters in the acts of annotation. We believe that introducing document parameters, time and the parameters of the creator of annotation into an annotation process can be a dependable source to know, who used a document, when a document was used and for what a document was used for. Beyond document tracking, our model can be used in not only for selective dissemination of information but for recommendation services. AMIEDoT can also be used for information sharing and information reuse.

**Categories and subject descriptors**

Information systems design, Document tracking, routing, recommenders, Information retrieval, filtering, and extraction

**Keywords**

Information research, model, document, user, time, annotation

**General terms**

Management, design, documentation

## 1. Introduction

Annotation has been a very useful tool in transmitting ideas from man to man. Not only that an annotation convey the thoughts of an initial user of a document to another user of the same document, it testify to the use of that document. The importance of annotation as a tool in information management can be seen with its popularity. Many text processors like Microsoft word, Adobe Acrobat and the like integrate features that enable users to annotate electronic documents. We believe that electronic annotation made by different users should not only be restricted to interpretation of the content of document(s); annotation tools can be designed to assist in recommendation service, information management and document tracking. It was on this basis that an annotation model AMIE was conceived.

## 2. Background

Annotation can be perceived from different perspectives and can assume different forms but for our study, we will define it as an action and an entity. From the perspective of an action, annotation can be defined as an act of interpreting a document. The interpretation is of a specific context and is expressed on the document. The interpretation can be made by the producer of the document or another person. It should be noted that when a document author makes annotations on his own document, he is seen at that moment as a reader of that document and not an author. Considering annotation as an entity, we define it as written, oral or graphic information usually attached to a host document meant to attest to the use of a document, for evaluation or interpretation of a document. Our study here we dwell on these two definitions of annotation interchangeably.

Electronic annotation can not take place until after the document has been made available to its audience. Every annotation on incomplete document is considered as part of the initial document. This is important as we apply annotation to published work. Annotations will normally take a different form and different look as compared to the original document. The difference in look may be noticeable in form of character used, font, style, color or additional signs and images that is not characteristic of the original document. The common intercept between annotation and the original document is the medium of transmission.

## 3. Constituents of an annotation

An annotation is essentially consisting of three main components; the annotator (person making the annotation), the document being annotated and the resulting annotation itself. We will not give attention to the annotator in this study because our concern here is not on user modeling or profiling. A document is defined in a general form as a trace of human activities [15]. A trace of human activities can include archaeological artifacts, buildings, cinema, books and monuments. In another word, an archaeological artifact is a document as much as a building. Though our finding in most of the cases is applicable to documents of various types, attention is given to written documents.

A document essentially contain information meant for interpretation (read, viewed, heard, perceived) by a certain group of people. The audience may or may not be pre-determined. It is therefore imperative that a document be made available to its potential audience. A document itself may be in oral, graphics or text form. It may be tangible or intangible.

Annotation can not take place until after the document has been completed. An annotation is not a property of a document. For instance, a plate number of a vehicle is not an annotation though it is attached to a vehicle. This is because, we consider a plate number as a property of a vehicle. A vehicle is a complete entity with a plate number. Annotations will normally take a different form and look with respect to the original document. The different in look may be noticeable in form of character used, font, style, color or additional signs and images that do not form part of the original document.

From our study of the literature on annotation, we were able to identify the following reasons why annotations are performed:

- Add an explanation to a document section (definitions, examples, references, etc.)
- Provides a means of evaluating a document (relevance of a document by providing a global point of view or a detailed evaluation criteria)
- Associate specific interpretation to a section of a document or to the document in its entirety, by giving additional attribute to the document with an associated value
- It could be used as a medium of information sharing,
- It may serve as a means of sieving information.
- Means of interpretation of document,
- It is a means of creating a forum for independent view of document,
- Facilitate critical reasoning,
- Permit the user to construct a personal representation of the document,
- They can attest to a witness of personal commitment by a reader to a document ,
- Permit monitoring trace of document use,

It should be noted that annotation does not result in the modification of the initial document. It may however constitute a new document for the reader. This point is essential in the sense that the author's copyright is protected.

According to Bringay et al [5]. annotation helps in the legibility of information. Annotation may at one time make the document legible but may also hinder the legibility of the same document at another time. It does not necessarily aid in making the information clear but gives a specific interpretation to the information contained therein.

Annotations are performed by users who have the intention of storing their point of view for future reuse. Among the users (or readers) are students, researchers, lecturers, or the general public. Annotations can be made manually. For example, stickers or post-it can be scotched at specific pages of a book. Specific colours may be used to underline a section of a document in order to specify the importance of that section. It could also be in form of underlining. Text grouping with the use of brackets or braces is sometimes used to annotate. It may also be in form of passage or paragraph numbering.

With electronic software, it is possible to create manual annotations and also store them for future and more elaborate use.

A person making an annotation has an objective in mind. He is making annotation to achieve among others reasons: He could be describing (summarizing) or evaluating (analyzing) an informative resources based on standard criteria.

## 4. Existing models

The basic objective of annotation conception is to provide for additional set of information that was not specified by the initial author of the document. This information is saved to the original document and referenced by a link. The goal of annotation is to allow addition to existing resources by individuals who normally will not have direct control on the original document.

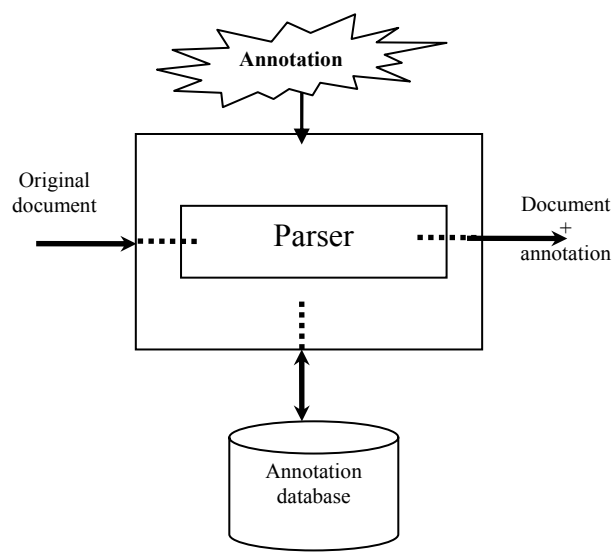

Figure 1: Architecture of generalized annotation system

Table 1: Document and user tracking based on log on documents (annotations)

| | Document and user tracking | Fixed parameters | | | Representation | Value range [u=user,d=doc,t=time] |
|---|---|---|---|---|---|---|
| | | User | Doc | Time | | |
| 1 | Annotations of all users of all document | | | | $\iiint dUdDdT$ | $[0 \leq u < \infty], [0 \leq d < \infty], [0 \leq t < \infty]$ |
| 2 | Annotations of all users of all documents at a specific time | | | X | $T\iint dUdD$ | $[t=1], [0 \leq u < \infty], [0 \leq d < \infty]$ |
| 3 | Annotations of all users of a document | | X | | $D\iint dUdT$ | $[d=1][0 \leq u < \infty], [0 \leq t < \infty]$ |
| 4 | Annotations of all users of a document at a time specific | | X | X | $DT\int dU$ | $[d=1][t=1], [0 \leq u < \infty]$ |
| 5 | Annotations on all documents used by a user all time | X | | | $U\iint dDdT$ | $[u=1][0 \leq u < \infty], [0 \leq d < \infty]$ |
| 6 | Annotations on all documents by a user at a specific time | X | | X | $UT\int dD$ | $[u=1][t=1] [0 \leq d < \infty]$ |
| 7 | Annotations by a user on a document | X | X | | $UD\int dT$ | $[u=1][d=1][0 \leq t < \infty]$ |
| 8 | Annotation on a document by a user at a specific time | X | X | X | $UDT$ | $[u=1],[d=1],[t=1]$ |

We can explain most models of annotation with figure 1. A document is sent to a parser with an annotation originating from the user of the system. The parser is considered as the motor of the system. How the annotation is made and to what part of document the annotation is addressed is what makes the difference. Generally annotation is added to the document based on a specific model. An annotation with the original document is created and returned to the user of the system. This created annotation is generally in form of the original document with a link (visible or invisible) pointing to the location of resulting annotation stored in an annotation database. The location of associated annotation database is also based on several factors depending on the level of security consideration. Some annotations are stored on the application server, which demand high security considerations. This type of system enhances optimal sharing of information between users but limits privacy. Some other annotation databases are stored on local machine. In this case, it enhances higher security but limit sharing of resources. A compromise between these two types of systems is the use of proxy-storage server. In this case another machine is situated in between the application server and the local machine to store annotations.

Several annotations systems were developed along the line of the structure of the document or based on the organization of the resulting annotation [1].[11]. Prominent among annotation system aimed at the structure of the document are annotation of type structuring and annotation for type classification. We can also consider annotations based on the methodology used in creating the annotations. Some are automatic, some semi-automatic and others are manual.

Annotation systems based on the structure of the document are concerned with the structural relationship between resulting annotation and the elements of the document annotated. Several annotation systems on the internet were conceived along this line. Example includes "annotation engine", Hylight, AMAYA, YAWAS [9]. and CritLink [20]. Annotation tools such as the one in Microsoft word is of the type structuring. Some annotation tools were conceived to classify documents [6].[11]. The inside structure of documents are not addressed but the general concept or interpretation of the entire document as a whole. An example of annotation for classification is Furl (http://www.furl.net).

Annotations tools based on the methodology of creation generally give rise to semantic annotation, ontological annotation or linguistic type of annotation. Some annotations tools were considered as functional [13]. They can still be seen as either based on the organization of the resulting annotation or based on the structure of the document.

## 5. Our approach

Our approach is from the perspective that annotation attest to the use of a document. The annotation may not necessarily be placed on the document but kept in a separate database and a link provided between annotation database and the bibliographic references. The objective of providing the link is so that a base for document tracking and document management may be created. Specifically, we want to be able to analyze the feedback made inform of annotation. These feedback will provide information like when was a document used? Who used a particular document? For what was a particular document used?

Among likely-hoods, we observed that (a) dissimilar document can be used differently by one or more users. (b) two or more individuals will not make use of the same document the same way (c) the same user may not use the same document the same way at different time. A document can be used several times by a particular individual. We presume that one document can not be used by two users at a time. One user may use a document different compared to another user. A particular user may use the same document differently given a time frame. In applying annotation tool into document tracking and recommendation service, we are considering document use (annotation) in terms of users, documents and time. Series of (document use) annotations over time on one or more documents, by one or more users can be used to evaluate the use of document and interest of individuals.

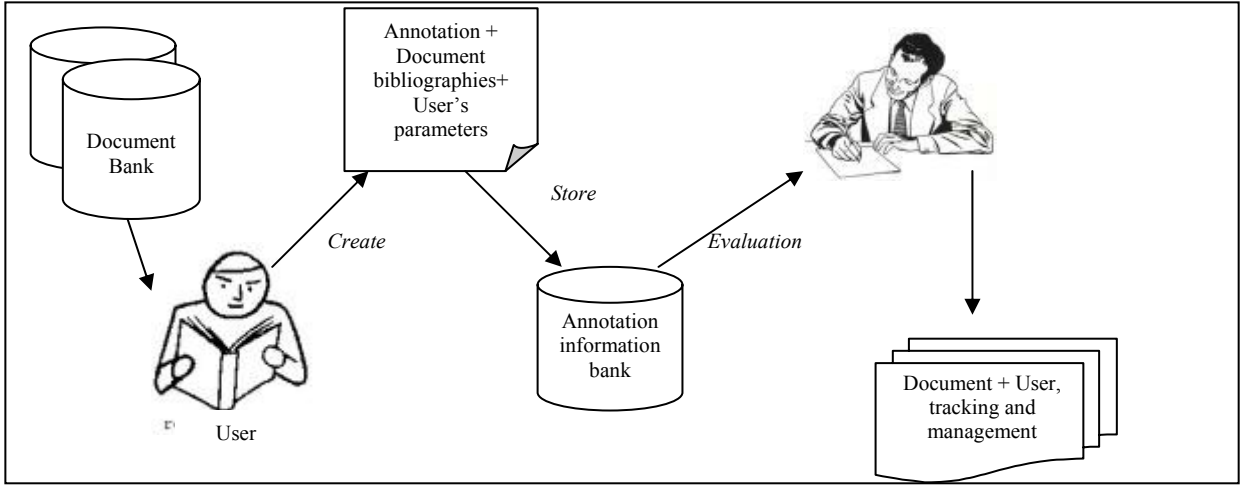

Figure 2: Application of AMIEDoT model for document and user tracking

An annotation or a set of annotation can be represented as

$$\iiint_x dU dT dD$$

Where $dU$ is the variation in users, $dT$ and $dD$ are variations in time and documents parameters respectively. Specifically, we are signifying that annotation can be seen as a function of user (U), time (T) and document (D).

One or more of these parameters can be kept constant while the other varied. The three parameters when kept constant refer to a single case of an annotation. In the case where all these vary, it imply every possible annotation on a set of documents of interest.

We can be interested in the document use made by a particular user on a particular document over time. The objective of this may be to see his reaction or the user's disposition to an event. We can represent this as

$$UD \int dT$$

We can represent this in a three dimensional graph with each of the parameters in X, Y and Z axis respectively or with a table as in Table 1. We used the word annotation in this table to include the log of document used. This is because some document may not necessarily be annotated.

## 6. Specifications in AMIEDoT model

Our model consists of four main entities considered as the core of the model: (a) the user who is also the annotator, (b) the document in question, (c) annotation transaction and (d) the process of annotation creation. We address the model from these four perspectives. Each of these parts has its characteristics and properties. We attempt to describe each.

**a. The user is the annotator**

The user information is generally available when a user signifies his intention in the use of the library. They are normally stored differently from library databases.

The user is identified with the following parameters.

- Annotator's reference (*this is a unique reference that is used to identify a user*).
- identity ( *identity of the document user*)
  - His name (*first name and last name*)
  - Email address
  - postal address (*or sectional address*)
  - region
  - age-group
  - country
  - social class
  - area of activity (*teaching, research, student etc*)
- session (*session is used to identify user's activities in the process with date and time*)

**b. Document**

The document consulted is paramount in any annotation process
- document title (*original title of document*)
- descriptors and keywords (*descriptors are words used to describe the document*)
- authors ( *are the producers of the document, their names and surnames*)

- date of publication of document
- format of document (PDF, word, html etc)
- abstract / résumé

**c. Annotation transaction (context of annotation stored in a storage)**

This is meant to store the session of user every time the system is consulted.
- approach {the type of annotation i.e. follow up or new annotation}
- context reference
- session reference (date/time)
- implicit parameters of the user
  - Length on system
  - Documents consulted
- why was the document consulted?
  - Leisure consultation
  - Knowledge acquisition
  - Accidental consultation
  - Academic reading
  - Research reference
  - To answer a question
  - Historic reference
  - Internet link
  - Other reasons
- explicit parameters of the users
  - user name

**d. Annotation creation**
- reference (*is the reference, or code for future reference*)
- type (*the type of annotation used*)
  » marking,
  » typographic
    ▪ italics, underlining ...
  » reformatting of text using brackets and braces,
  » passage numbering,
  » text
    ▪ in margin, footnotes, endnotes, in the gutter, by icons),
  » icons
    ▪ stars, question marks, exclamation marks,…
  » symbols
    ▪ to describe associations, relations between words..
- annotation location
    ▪ left margin, right margin, footer, header, gutter, outside document, end of document
- why annotating (objective of annotation)?
  » recapitulation, evaluation,
  » summary, raise a point,
  » classification, structuring,
  » differentiating, for information,
  » answer to a question,
  » illustration, extension of document,
  » clarify ambiguity of document

*Application of the model*

When a user request for a document from a document bank like in a library, usually, no record is made on the use of the document requested. We care considering the case where every user of every document in a library is recorded in term of the use of document, the period when document was used and annotations on documents used. This can be achieved by use of filling a page questionnaire. The form can contain very few questions like (a) Why the document was consulted? (b) Why was annotation made? and (c) and free form comment of users (annotation). Other parameters of the model can be provided by a librarian. In some of the cases, users may not provide any particular comment. Our interest is not necessarily on the comment he made, but on his profile and the consultation on the document and for what the document is used for. Because, most library users may not be willing to make comments, or assessment on every document used, for a start, for a start, we can apply this model to borrowed documents. We are aware that there exist necessary bibliographic information on all documents in libraries and database of users. It is the merging of bibliographic records with user's identity in time with additional comments that will provide vital information for document and user tracking.

A part from the fact that information on user and on document are the core of our model, we can find out among others the following information using the model: (a) The most consulted document (b) The most frequent objective of document consultation (c) the frequency of users in the library (d) Which user is in what domain? (e) Which user is related to another user from the perspective of the documents they consult? (f) What is the view of a user (or a group of user) with respect to a discipline - annotation of a user (group of users) with respect to documents in a discipline? (g) What social class consults most frequently? (h) What type of document is good for what user?

From the annotation made on document consulted, some of the information that can de derived include (a) The general interest of a user (or group of users to) (b) Type of document most interesting to a user (or a group of user)? (c) The most important objectives for making annotations? (d) The trend or general perception of a document or group of documents? Some of these analysis may be useful in classification or reclassification of document. It can even be used in associating key words and descriptions to document.

## 7. Perspective

The problem that has not been fully considered includes, how do we reconcile the changes that may exist in document with time in respect with usage? We are concerned with changes initiated by the author of the initial document. Of course, we can assume a different status for the "new document". The work on reference-based version model [7]. offer a great potential. It may be very interesting to see not just the use of document with time but the variation in document as well. A user may make use of a document differently if there are variations in the source document.

Our analysis does not consider intra-parameter considerations. For example, we may be interested in seen the variations in social class of users with type of annotation or document descriptions. These and other possible analysis is left to the discretion of the users.

## 8. Conclusion

From our studies, we have shown that annotation can be very useful in document tracking and user analysis. Annotation has been viewed as a function of its maker (the annotator), the document been annotated and the time of annotation. These three parameters are very important in its application to document/user tracking and management.

## REFERENCES

[1] D. Bargeron, A. Gupta and A.J. Brush Bernheim., A Common Annotation Framework, Technical Report MSR-TR-2001-108, Microsoft Research, Microsoft Corporation, One Microsoft Way, Redmond, WA 98052, 2001

[2] A.J. Bernheim Brush, Annotating Digital Documents for Asynchronous Collaboration, Department of Computer Science and Engineering, University of Washington, Technical Report 02-09-02, September 2002, 2002

[3] D. Bueno and David A., METIORE: A Personalized Information Retrieval System Source Proceedings of the *8th International Conference on User Modeling 2001*, in Lecture Notes In Computer Science; Vol. 2109, Pages: 168 - 177 , 2001

[4] N. Bouaka and David A., Modèle pour l'Explicitation d'un Problème Décisionnel: Un outil d'aide à la décision dans un contexte d'intelligence économique. *in Conférence "Intelligence Economique : Recherches et Applications"*, Nancy : 14-15 avril 2003. , 2003

[5] S. Bringay, C. Barry and J. Charlet, Les documents et les annotations du dossier patient hospitalier, « *Information-Interaction-Intelligence* » Volume 4, n°1, Juillet 2004, 2004

[6] R. Casallas-Gutierrez, Objets historiques et annotations pour les environnements logiciels, Thèse de Doctorat à l'Université Joseph Fourier Grenoble I, May 1996. , 1996

[7] S.Y. Chien, V. Tsotras and J. Zaniolo, Carlo Efficient Management of Multiversion Documents by Object Referencing, *Proceedings of the 27th VLDB Conference*, Roma, Italy, 2001

[8] A. David, D. Bueno, and P. Kislin, Case-Based Reasoning, User model and IRS. In *The 5th World Multi-Conference on Systemics, Cybernetics and Informatics - SCI'2001. International Institute of Informatics and Systemics (IIIS). (Orlando, USA)* , 2001

[9] L. Denoue, L. Vignollet, An annotation tool for Web browsers and its applications to information retrieval. *Proceedings of RIAO 2000* , 2000

[10] S. Goria and P. Geffroy, Le modèle MIRABEL : un guide pour aider à questionner les Problématiques de Recherche d'Informations.*In Veille Stratégique Scientifique et Technologique - VSST'2004 (Toulouse).* 2004. 5p. , 2004

[11] R. M. Heck, S. M. Luebke, and C.H. Obermark, A Survey of Web Annotation Systems, Internal report, Department of Mathematics and Computer Science, Grinnell College, USA, 1999

[12] P. Kislin and A. David, De la caractérisation de l'espace-problème décisionnel à l'élaboration des éléments de solution en recherche d'information dans un contexte d'Intelligence Economique : le modèle WISP, *in Conférence "Intelligence Economique : Recherches et Applications"*, Nancy : 14-15 avril 2003, 2003.

[13] B.K. Liisa, M. W. Gary, B. FZ Lang B. and G. Burgern, AutoFACT : Automatic Functional Annotation and Classification Tool, BMC Bioinformatic, 6:151, 2005

[14] H. Martre, « Intelligence économique et stratégie des entreprises », *Rapport du commissariat Général au Plan*, Paris, La documentation Française, pp 17,18, 1994

[15] Y. Prie, « Modélisation de documents audiovisuels en Strates Interconnectées par les Annotations pour l'exploitation contextuelle » Thèse de Doctorat à l'Université Claude Bernard Lyon1, France, 1999.

[16] A.B.C. Robert, Représentation des activités du veilleur en contexte de l'intelligence économique, DEA en Sciences de l'information et de la Communication, Université Nancy 2, Université de Metz, Octobre 2003, 2003

[17] M. A. Schickler, M.S. Mazer and C. Brooks, Pan-Browser Support for Annotations and Other Meta-Information on the World Wide Web, *Fifth International World Wide Web Conference* May 6-10, 1996 , Paris, France, 1996

[18] S. Sidhom,"*Plate-forme d'analyse morpho-syntaxique pour l'indexation automatique et la recherché d'information: de l'écrit vers la gestion des connaissances.*", Thèse de Doctorat à l'Université Claude Bernard Lyon1, France, Mars 2002. , 2002

[19] D.V. Sreenath, W. Grosky and F. Andres, *Intelligent Virtual Worlds: Technologies and Applications in Distributed Virtual Environments*, chapter Metadata-Mediated Browsing and Retrieval in a Cultural Heritage Image Collection. World Scientific Publishing Company, Singapore, 2002

[20] K.P. Yee, CritLink: Advanced Hyperlinks Enable Public Annotation on the Web. Demonstration abstract. ACM Conference on Computer-Supported Co-operative Work, , 2002